# Demonstration of a passive photon-atom swap gate

Orel Bechler[1†], Adrien Borne[1†], Serge Rosenblum[12†], Gabriel Guendelman[1], Ori Ezrah Mor[1], Moran Netser[1], Tal Ohana[1], Ziv Aqua[1], Niv Drucker[1], Ran Finkelstein[1], Yulia Lovsky[1], Rachel Bruch[1], Doron Gurovich[1], Ehud Shafir[1] and Barak Dayan[1*]

**Deterministic quantum interactions between single photons and single quantum emitters are a vital building block towards the distribution of quantum information between remote systems[1-4]. Deterministic photon-atom state transfer has been demonstrated by using protocols that include active feedback or synchronized control pulses[5-10]. Here we demonstrate a completely passive swap gate between the states of a single photon and a single atom. The underlying mechanism is single-photon Raman interaction (SPRINT)[11-15] – an interference-based effect in which a photonic qubit deterministically controls the state of a material qubit encoded in the two ground states of a $\Lambda$ system, and vice versa. Using a nanofiber-coupled microsphere resonator coupled to single Rb atoms we swap a photonic qubit into the atom and back, demonstrating nonclassical fidelities in both directions. Requiring no control fields or feedback protocol, the gate takes place automatically at the timescale of the atom's cavity-enhanced spontaneous emission time. Applicable to any waveguide-coupled $\Lambda$ system, this scheme provides a versatile building block for the modular scaling up of quantum information processing systems.**

[1] AMOS and Department of Chemical Physics, Weizmann Institute of Science, Rehovot 76100, Israel.
[2] Current address: Departments of Physics and Applied Physics, Yale University, New Haven, CT 06520, USA.
[*] e-mail: barak.dayan@weizmann.ac.il
[†] These authors contributed equally to this work.

While teleportation protocols[16-18] provide one possible route for linking separate quantum modules, considerable effort is currently invested towards the realization of a direct state transfer between photonic and material qubits, especially within the field of cavity quantum electrodynamics[19,20]. Already at moderate fidelities and efficiencies[21], such photonic links enable scalable architectures for distributed quantum information processing based on interconnected compact modules[1-4].

Deterministic two-way transfer between photonic and material qubits had so far been demonstrated only with active protocols that require auxiliary control fields – from stimulated Raman adiabatic passage[6,7], to the Duan and Kimble protocol[5], which uses classical laser pulses to manipulate and measure the state of the atom[8]. These approaches were applied to demonstrate both photon-atom state transfer[7,10], and photon-atom and photon-photon gates[8,9,22]. In this work we introduce an alternative approach for deterministic quantum interactions between photonic and material qubits, one that is completely passive and can be used both for quantum state transfer and for universal quantum gates. The underlying mechanism, SPRINT, harnesses quantum interference to "force" a single Λ-type quantum emitter to end up in one of its two ground states depending on the input mode of a single photon. Initially considered in ref. 23 and analyzed in a series of theoretical works[15,24-27], SPRINT was first realized experimentally to create an all-optical single-photon switch[11]. It was then harnessed for deterministic extraction of a single photon from optical pulses[13]. The effect was also demonstrated in superconducting circuits, where it enabled frequency conversion of microwave fields[12] and highly efficient detection of single microwave photons[14]. The configuration that leads to SPRINT includes a Λ-system in which each of its two transitions is coupled, with the same cooperativity $C$, to a different mode of an optical waveguide. As depicted in Fig. 1a, assuming the system is initiated at ground state $|\downarrow_z\rangle_a$, and that losses and coupling to free-space modes (at rate $\gamma$) are much smaller than the coupling rate $\Gamma$ to the waveguide modes (i.e. $C = \Gamma/\gamma \gg 1$), the radiation from the system interferes destructively with any incoming light in mode $\hat{a}$, eliminating the outgoing field in this mode.

Any incoming photon is therefore diverted to mode $\hat{b}$, leading to a Raman transition of the atom to state $|\uparrow_z\rangle_a$. If the system is initiated instead in ground state $|\uparrow_z\rangle_a$ (Fig. 1b), it does not interact with the photon, which therefore continues undisturbed in mode $\hat{a}$. Evidently, the photon and the Λ-system play symmetric roles: the incoming mode of the photon dictates the final state of the Λ-system, and the initial state of the Λ-system dictates the output mode of the photon. As SPRINT is coherent, this description holds for superposition states as well. This means that it essentially performs as a quantum swap gate between the photonic qubit (encoded in a superposition of the two input modes) and the material qubit (encoded in its two ground states).

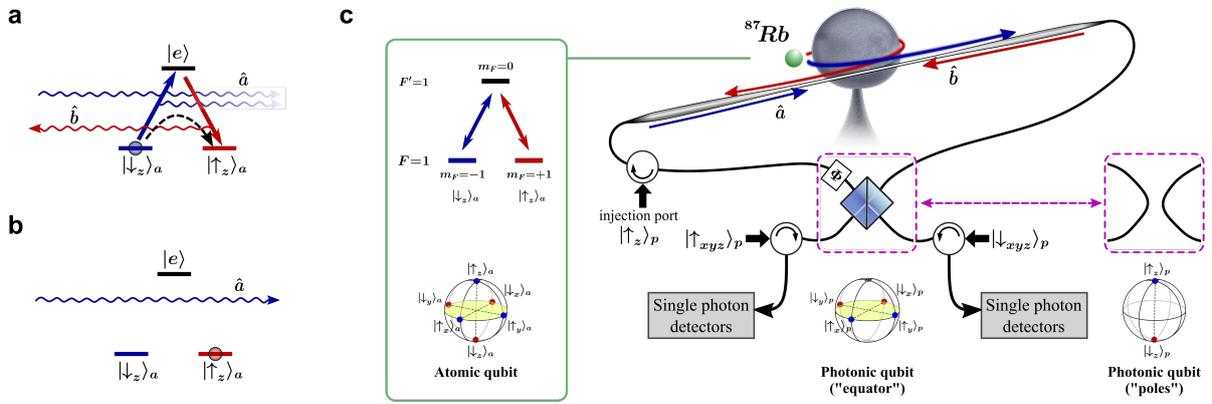

**Figure 1. The single-photon Raman interaction (SPRINT) scheme and experimental apparatus**. **a,** A material Λ-system in which each transition is coupled to a different single optical mode ($\hat{a}$, $\hat{b}$). Incoming light from the left ($\hat{a}$) interacts with the system, initially in ground state $|\downarrow_z\rangle_a$ (blue). Destructive interference (blue gradient fade) in this mode forces the system to radiate only to the left ($\hat{b}$), thereby toggling the system to the right ground state $|\uparrow_z\rangle_a$ (red). **b,** In its toggled state, the system is now transparent to light incoming from the left. As evident, the input photonic mode determines the resulting state of the material system and vice versa. **c,** A single $^{87}$Rb atom (green sphere) interacts with a TM mode of a microsphere resonator. The input-output interface is provided by an optical nanofiber in which mode $\hat{a}$ (blue) propagates from left to right, and mode $\hat{b}$ (red) from right to left. Optical circulators (see Methods) are used to properly guide light into and out of the tapered fiber. Four input directions are available: $|\uparrow_z\rangle_p$, $|\downarrow_z\rangle_p$ (corresponding to the "pole" states of the photonic qubit Bloch sphere) and their superposition ("equator") states, accessible using a 50/50 beam splitter combined with a Pockels cell to control the phase (Φ). Outputs are split using fiber couplers to ten single-photon counting modules (SPCMs), five on each output port (not shown). The "injection" port on the left side enables probing the atom by sending a photon with a fixed photonic state $|\uparrow_z\rangle_p$.

Our experimental implementation consists of a single $^{87}$Rb atom coupled to an ultra-high quality microsphere resonator, which is temperature-tuned to be resonant with the $F = 1 \to F' = 1$ transition of the $^{87}$Rb D$_2$ line. The atoms originate in a magneto-optical trap, where they are initially trapped, cooled to ~7 µK and finally dropped into the vicinity of the microsphere placed ~7 mm below the trap. Falling through the evanescent field of one of the resonator's transverse-magnetic (TM) whispering-gallery-modes (WGMs), atoms are detected using a fast series of pulses before and after the experiment is carried out, ensuring they reside within the optical mode throughout the measurement (see Methods). Light is evanescently coupled into and out of the resonator using a tapered nanofiber (Fig. 1c). The optical setup has two possible configurations – with and without a 50/50 beam splitter. Without the beam splitter, the two input/output ports couple light to opposite directions of the tapered fiber, corresponding to the "poles" of the photonic qubit Bloch sphere, $|\uparrow_z\rangle_p$ and $|\downarrow_z\rangle_p$. With the beam splitter, along with an electro-optic phase modulator (Pockels cell), the two input/output ports correspond to the "equator" states on the photonic Bloch sphere - the superpositions $(|\downarrow_z\rangle_p + e^{i\Phi}|\uparrow_z\rangle_p)/\sqrt{2}$ and $(|\downarrow_z\rangle_p - e^{i\Phi}|\uparrow_z\rangle_p)/\sqrt{2}$.

Conveniently, the two counter-propagating TM modes of the microsphere exhibit nearly opposite circular polarizations[28,29]. This means that the photonic mode $\hat{a} = |\uparrow_z\rangle_p$ drives mostly the atomic $m_F = -1 \to m_{F'} = 0$ transition, and the mode $\hat{b} = |\downarrow_z\rangle_p$ drives mostly the $m_F = +1 \to m_{F'} = 0$ transition, with little crosstalk (~4.5%). This system therefore approximates well the desired SPRINT configuration, which creates a one-to-one correspondence between the photonic Bloch sphere ($|\uparrow_z\rangle_p$, $|\downarrow_z\rangle_p$ and their superpositions) to the atomic Bloch sphere (the states $|\uparrow_z\rangle_a = |F = 1, m_F = +1\rangle$, $|\downarrow_z\rangle_a = |F = 1, m_F = -1\rangle$ and their superpositions).

Given the intrinsic cavity loss rate $\kappa_i/2\pi = 6$ MHz of our chosen WGM, we tune the fiber-cavity coupling rate to be $\kappa_{ex}/2\pi = 60$ MHz in order to optimize the interference and maximize our SPRINT performance[15]. This sets the linear loss of the bare cavity to ~30% on resonance. With a coherent atom-cavity coupling rate of $g/2\pi = 27$ MHz and atomic free-space amplitude decay rate of $\gamma/2\pi = 3$ MHz, our system is in the fast-cavity regime for which $\kappa_{ex} \gg g \gg \kappa_i, \gamma$. The resulting cavity-enhanced spontaneous emission rate in both fiber directions is $\Gamma/2\pi \sim 2g^2/(\kappa_{ex} + \kappa_i) \sim 22$ MHz (see Methods). With these parameters, the efficiency of the system remains close to ~70% also when coupled to the atom[15].

Demonstrating the feasibility of using SPRINT for photonic quantum communication links requires assessing its performance in mapping an atomic qubit onto an outgoing photon and vice versa. To demonstrate the first part of this twofold task, we first prepare the atom in one of the six cardinal points on the Bloch sphere (namely $|\uparrow_{xyz}\rangle_a$ and $|\downarrow_{xyz}\rangle_a$, see Methods). Next, a weak probe pulse (~0.05 photons on average) is sent to the atom. If a photon is present in the pulse, it should swap its state with that of the atom. Accordingly, the outgoing photon will ideally propagate in a direction dictated only by the atomic state, whereas the direction of the incoming probe photon should be irrelevant. In order to determine the fidelity of the output state with the initial atomic state, we analyze it in the basis along which the atomic qubit was prepared.

The results, shown in Fig. 2, are presented for two variations of the atom-to-photon swap process. In the first realization, we send the probe photon along the same axis in which the atom was prepared. This in fact demonstrates the coherence of the SPRINT mechanism, as it works equally for the cardinal states in all the bases: the probe photon is mostly transmitted, i.e. stays in the same mode, if it matches the state of the atomic qubit (T~86%, R~14%), and mostly reflected, i.e. switches to the other mode, if it is in the orthogonal state

(T~32%, R~68%, see Fig. 2a). These measurements correspond to an average atom-to-photon swap fidelity of $\bar{\mathcal{F}}_{AP} = 77.2(\pm 2.1)\%$.

However, the fact that the axis of the probe photon matches that of the prepared atomic state can potentially bias the atom-to-photon swap process, even though ideally it should affect only the resulting atomic state. To prevent such a bias we also measured the fidelity of the atom-to-photon swap process by sending the probe pulse from a fixed direction ($|\uparrow_z\rangle_p$ through the "injection" port in Fig. 1c), regardless of the preparation basis of the atomic state. In this case the measured average fidelity was $\bar{\mathcal{F}}_{AP} = 74.7(\pm 1.7)\%$ (Fig.2b).

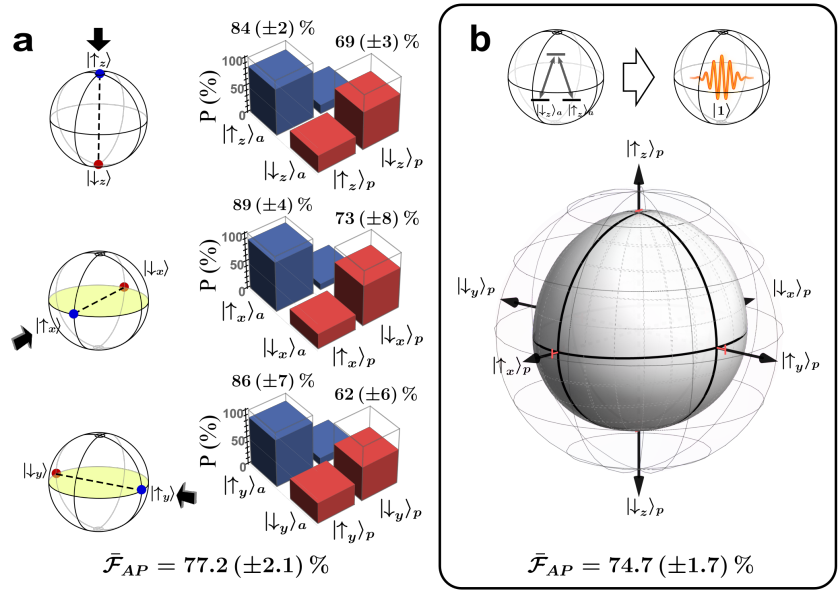

**Figure 2. Experimental results for the atom-to-photon swap process in two configurations. a,** The measured performance of SPRINT for initial atomic states on all the three axes of the Bloch sphere. The atomic state in each of the cases was prepared either at $|\downarrow\rangle_a$ or $|\uparrow\rangle_a$ on a chosen axis. A probe pulse was then sent from the $|\uparrow\rangle_p$ direction on the same axis (black arrows). Bar graphs represent obtained probability for $|\uparrow\rangle_p$ or $|\downarrow\rangle_p$ (corresponding to transmission or reflection, respectively) at the output for each prepared atomic state (blue or red bars), empty black boxes signify unit fidelity. The presented results correspond to an average fidelity of $\bar{\mathcal{F}}_{AP} = 77.2(\pm 2.1)\%$. **b,** Measured fidelity of the resulting photonic state for each of the six input atomic cardinal states, here with the probe photon always sent from port $|\uparrow_z\rangle_p$. The average fidelity is $\bar{\mathcal{F}}_{AP} = 74.7(\pm 1.7)\%$.

In order to evaluate the fidelity of the swap operation in the other direction, i.e. the photon-to-atom state transfer, we conducted a double, swap-in-swap-out experiment by sending the six cardinal photonic input states, storing them for 300 ns in the atom, and then reading them out by sending a probe photon from port $|\uparrow_z\rangle_p$. The "write" pulse which performed the first swap operation that mapped the photonic state into the atom had ~0.8 photons on average, and the probe pulse that swapped the atomic state back to a photon had ~0.05 photons on average. Both pulses were approximately 50 ns long. The state of the output photon was then analyzed on the same axis as the input.

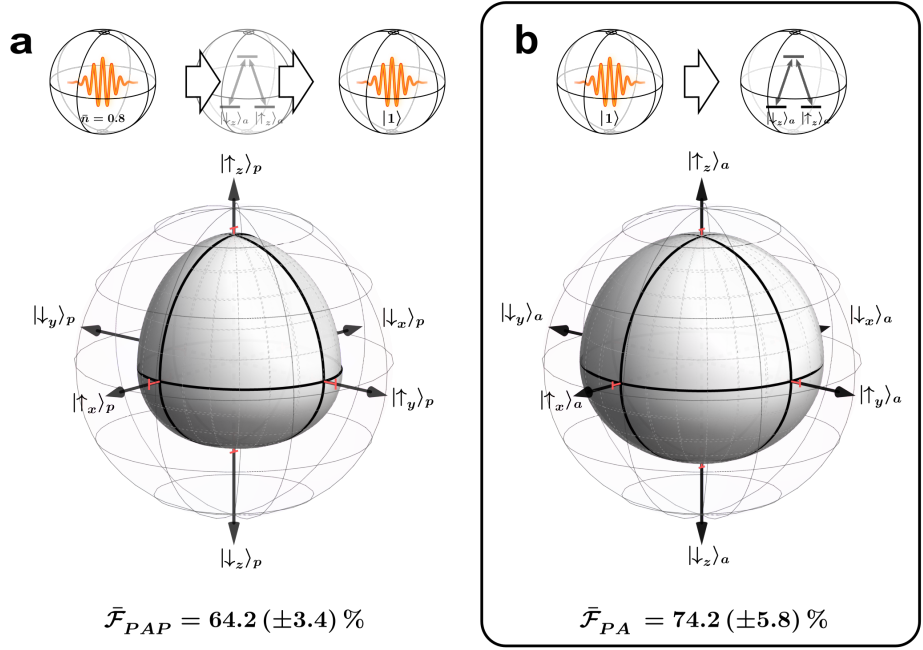

$\bar{\mathcal{F}}_{PAP} = 64.2\,(\pm 3.4)\,\%$  $\bar{\mathcal{F}}_{PA} = 74.2\,(\pm 5.8)\,\%$

**Figure 3. Measurement of the photon-to-atom swap fidelity. a,** Measured fidelity of the resulting photonic state for each of the six input photonic cardinal states for the photon-atom-photon double-swap process. A coherent pulse with an average of 0.8 photons "writes" a photonic qubit on an atom using SPRINT. The state is then extracted as a single photon (by sending a probe pulse from port $|\uparrow_z\rangle_p$) and analyzed in the write basis. The average fidelity measured for the entire process is 64.2($\pm$3.4)%. **b,** Inferred fidelity of the resulting atomic state for each of the six input photonic cardinal states for the photon-atom swap process. Results assume a perfect single-photon input qubit and perfect atomic preparation at $|\uparrow_{xyz}\rangle_a$. The average fidelity results in 74.2($\pm$5.8)%.

The results, shown in Fig. 3a, yield an average measured fidelity of $\bar{\mathcal{F}}_{PAP} = 64.2(\pm 3.4)\%$. These measurements were heralded on the detection of a single photon in the writing pulse, to rule out the ~45% probability of vacuum in the writing pulse. Without heralding the measured average fidelity was $60.0(\pm 1.3)\%$ (see Methods for further details). Due to the Poissonian statistics of the writing pulse, the fidelity obtained from both measurements (heralded and unheralded) do not directly represent the fidelity for the double-swap operation for an ideal input state of a single photon (see Methods for comparison with the classical threshold). Yet by analyzing these results together with the measured results of the atom-to-photon swap process (Fig. 2b), one can infer the fidelity of the photon-to-atom swap process for a perfect single-photon input (see Methods). The resulting average fidelity is $\bar{\mathcal{F}}_{PA} = 74.2(\pm 5.8)\%$ (Fig. 3b).

In conclusion, we have demonstrated a passive mechanism for bidirectional conversion between material and flying photonic qubits, which takes place on the time-scale of the Purcell-enhanced spontaneous emission time $\Gamma^{-1}$. Requiring no control fields, our SPRINT-based swap gate is suitable for scalable, 'digital-circuit-like' quantum networks, in which the output photonic qubit from one node can immediately serve as the input to the next one. The fidelity and efficiency of the swap operation in our current realization agree well with the theoretical prediction[15] and are limited mainly by the imperfectly circular polarization of the TM mode, position-dependent variations in $g$, and the intrinsic linear loss of the WGM. While all these parameters can be further optimized, the potential of this scheme is the fact that it can be applied to any material Λ-system coupled to a waveguide, and can further serve as a building block for universal quantum gates such as $\sqrt{SWAP}$[21] and C-Phase[30], thereby providing a versatile platform for quantum communication and distributed quantum information processing.

**Methods**

**Experimental sequence.** Acquisition of experimental data was performed by repeatedly (~1 Hz) releasing a cloud of ~90 million $^{87}$Rb atoms laser cooled to ~7 μK onto a silicon chip carrying WGM microsphere resonators, 7 mm below. Light was coupled to a chosen WGM of a single resonator by a tapered nanofiber (Fig. 1c). All the optical input/output ports of the nanofiber included 97:3 beam-splitters that functioned as circulators: attenuating the input weak coherent pulses and highly transmitting the output photons. At each drop, the single-photon counting modules (SPCMs) were triggered and photon detection events were recorded for 30 ms. During this time all the trapping beams were turned off and repeated sequences of pulses were sent in an interleaved pattern to the input ports of the nanofiber, both to detect the presence of an atom within the evanescent field of the WGM and to swap a photonic qubit into the atom or read it back (see below). The magneto-optical trap was then turned back on and the cloud regenerated and cooled again to ~7 μK using polarization-gradient cooling. Between each drop the WGM resonance was scanned and locked to the F=1 → F'=1 $^{87}$Rb D$_2$ transition, on resonance with the probe. Cavity locking was done by controlling the temperature of the chip using a combination of a thermo-electric cooling element along with free-space illumination of the microsphere using a few mW of a 532 nm laser.

**Atom detection and pulse sequence.** For detection of the atoms and carrying out the swap gates attenuated coherent pulses were sent to the nanofiber through two or three input ports, each corresponding to a different state on the photonic qubit sphere (Fig. 1c). These sequences consisted of four detection pulses, each ~10 ns long and with a mean photon number of ~1.2, coming alternately from opposite ports of the photonic Bloch sphere, preceding two weaker and longer pulses: "swap-in" and "swap-out" (~0.8, ~0.05 photons respectively, both ~50 ns), and finally four more pulses, identical to the initial detection pulses. The extremely low mean photon number in the swap-out pulse is meant to minimize the probability for the presence (and possible detection) of a second photon in the pulse, leading to failure of the readout process (the very rare probability for such an event was nonetheless taken into account in our analysis). All the detection pulses are meant to toggle the state of the atom by SPRINT, each toggle leading to reflection of the incoming photon with high probability, thereby indicating the presence of an atom in the mode. For the atom-to-photon swap experiments, the swap-in pulse served as a control pulse by conditioning the measurement on a reflection of a photon from this pulse, which indicates with high probability a successful swap operation and therefore preparation of the atom in the chosen state. The seventh pulse (the first of the last four detection pulses) was ~20% stronger and served as an erasure pulse, meant to re-initialize the state of the atom after the measurement, and was not used for detection, to prevent biasing of the atomic detection probability by the possible success or failure of the preceding swap operation. In the photon-to-atom swap experiments no conditioning was performed on the initial atomic state. Conditioning on the detection of a reflection event in at least three detection pulses, some before and some after the two swap pulses, indicates an atom was present within the cavity mode at probability that varied between 92.5% to 94.2%. The false detection probability was measured periodically by sending detection pulse sequences to the cavity after the atomic cloud had passed, before the next drop. This statistic was then used to correct our measured data. Overall, data was acquired from a total of ~86000 detected atoms in over 610,000 experimental cycles. All pulses were generated and shaped using an intensity electrooptic modulator (Photline NIR-MX800-LN-10) controlled by an arbitrary-waveform generator (Tektronix AWG7052). The pulses were then split into three acousto-optic modulators (AOMs), used for channeling pulses into the various input ports to the nanofiber (Fig 1c) with >30 dB extinction ratio.

**Data analysis.** Data was collected using a total of ten SPCMs (six Excelitas SPCM-AQRH-14-FC and four additional modules in a Perkin-Elmer SPCM-AQ4C-FC array), five on each side of our apparatus (Fig. 1c), and recorded using a photon correlator (Becker & Hickl DPC-230). In order to nullify any SPCM after-pulsing effects in our analysis, we disregard any detections which were separated by less than 200 ns on a given module. Accordingly we designed our pulse sequences so that the two swap pulses begin >200 ns after the end of the detection pulses and are spaced by more than 200 ns (tail-to-tail) from each other as well (>300ns peak-to-peak).

**Linear loss and single photon detection efficiencies.** The transmission of the bare tapered fiber was 90%, the optical-path efficiency starting from the output of the fiber to the detectors was 61.5% and the detection efficiency of the single-photon counting modules ranges between 55% to 60%. The linear cavity loss of ~30% and atomic spontaneous emission to free space were the only optical losses that were considered inherent to this realization of the swap gate. During some of our "pole" measurements, there was an additional asymmetric loss of 20% - 34% caused by a defect that formed on our nanofiber to one side of the cavity. By continuously monitoring the residual reflection from the empty cavity in both directions and comparing them to each other and to the overall transmission we were able to compensate for this loss and correct our results accordingly.

**Inferring photon-to-atom swap fidelity.** As described in the main text, using our measured experimental data one can numerically infer the average photon-to-atom swap fidelity. We represent each single-swap process as a probability table, their combination yielding the double-swap fidelity. We define $P_{j,i}$ as the probability to $j$ when expected to do $i$ and denote $i,j = \{t, nt\}$ for atomic state "toggle" or "no toggle" and $i,j = \{R, T\}$ for photonic "reflection" or "transmission", respectively. As a first step, we experimentally deduce $P_{R,R}$ and $P_{T,T}$ (from $\bar{\mathcal{F}}_{AP}$, Fig. 2b) and use them to generate all possible combinations for $P_{t,t}$ and $P_{nt,nt}$ and their return values $\bar{\mathcal{F}}'_{PAP}$. Each pair was given a weight to quantify its result's proximity to the actual measured value $\bar{\mathcal{F}}_{\text{PAP}}$ (Fig. 3a). The next step was to re-calculate the $\{P_{R,R}, P_{T,T}\}$ table using the $\{P_{t,t}, P_{nt,nt}\}$ table by applying the same method. The output was a new set of $P_{R,R}$ and $P_{T,T}$ values, each pair with its appropriate weight. These two steps were repeated until they converged onto two sets of pairs that can be used to calculate the average fidelity of the photon-to-atom ($\bar{\mathcal{F}}'_{\text{PA}}$) and atom-to-photon ($\bar{\mathcal{F}}'_{\text{AP}}$) swap processes, along with their respective errors. Performing this procedure with an atom initially populated at 50/50 and subsequently interacting with a series of detection and measurement pulses identical to our experimental sequence produces the desired average fidelity of our gate for an atom perfectly prepared at $|\uparrow_{\text{xyz}}\rangle_a$ and a true single-photon input. Throughout the calculation, realistic representation of pulses was carried out by assuming Poissonian photon-number distribution. Effectively, this meant applying the single-photon table according to the appropriate Poissonian probability for a single-photon, applying it twice according to the probability for two-photons, or not applying anything according to the probability for vacuum.

**False atomic preparation.** Using the calculations described in the previous section, backed by measurements gathered from our previous experiments[13], we can quantify the false atomic preparation probability in our system with high accuracy. As stated in the main text, this is mostly attributed to crosstalk between the not completely circular TM modes[29]. This results in a ~4.5% probability to record a reflection in our detectors although the atom decays to the wrong ground state. Our swap-out results take these pre-measurement preparation errors into account (see Extended Data Table 1 for a summary of all results).

**Classical threshold of the swap gate.** In order to ascertain the quantum nature of our swap gate, we must compare our measured average fidelity values to those obtainable in a classical system. Unlike a quantum process, a classical process initially involves projecting a general incoming quantum state onto a chosen axis. Assuming no prior knowledge of the incoming qubit axis, the highest achievable fidelity is therefore limited. Choosing the correct axis would project the state correctly, while the other two axes yield the correct state in only 50% of the cases. This sets the classical fidelity threshold for a single-photon[31] at 2/3. Our measured single-swap average fidelities in both directions (74.2%-77.2%) exceed this limit.

**Classical threshold of the swap-in-swap-out process.** The double-swap operation consists of two independent processes that do not share any knowledge of their axis operation. Therefore, each of the operations should be limited to 2/3. The result of the double process in the classical case is therefore expected to be the sum of the probability for a correct swap in both directions (4/9) and the probability for two wrong swaps, which also leads to the correct result in 1/9 of the cases. For a true single-photon input, the classical limit is therefore 5/9 ~ 56% (if the write and read processes are allowed to share information about the chosen axis the threshold remains 2/3). In the case of a coherent state with $n = 0.8$, one must take into account the Poissonian nature of the input pulse[6], which increases the classical threshold for the photon-to-atom swap to ~70% and the total threshold to ~57%. Our measured average fidelities in this case (64.2%) is above the classical limit, as well.

**Comparing heralded and unheralded results.** Results for the double-swap and photon-to-atom swap experiments can be analyzed for two possible modes of operation. One is heralding on the detection of a photon at the end of the writing stage, as discussed in the main text. This indicates both that the writing pulse had at least one photon, and that this photon was not lost after or (more importantly) before the interaction with the atom. Another mode of operation is the unheralded one, in which case the write pulse can sometimes be vacuum. As shown in Fig. 3a, the average measured fidelity in the heralded case is $\bar{\mathcal{F}}_{PAP} = 64.2(\pm3.4)\%$. Analysis for the unheralded case yields an average measured fidelity of

60.0($\pm$1.3)%. Beyond the fact that in the unheralded case the writing pulse sometimes (~45%) contains no photons at all, in both cases there is also probability that it had two photons or more. Inferring the fidelity of the photon-to-atom swap process for a perfect single-photon input, the resulting average fidelity (Fig. 3b) is $\bar{\mathcal{F}}_{PA} = 74.2(\pm5.8)$% in the heralded case, 69.9($\pm$2.4)% when unheralded. A difference of ~4% between these two numbers is indeed expected, resulting mostly from the 30% probability for linear loss of the write photon within the cavity, which in half of the cases occurs before it reaches the atom, and brings the expected photon-to-atom swap fidelity to that of vacuum, namely 50% (see Extended Data Table 1).

**atom-to-photon swap**

|  | same-axis probe | | fixed-direction probe | |
|---|---|---|---|---|
|  | $\vert\downarrow\rangle_p$ | $\vert\uparrow\rangle_p$ | $\vert\downarrow\rangle_p$ | $\vert\uparrow\rangle_p$ |
| poles | 70($\pm$3)% | 87($\pm$2)% | 70($\pm$4)% | 87($\pm$3)% |
| equator | 69($\pm$3)% | 91($\pm$2)% | 77($\pm$3)% | 80($\pm$3)% |

**double-swap**

|  | heralded | | unheralded | |
|---|---|---|---|---|
|  | $\vert\downarrow\rangle_p$ | $\vert\uparrow\rangle_p$ | $\vert\downarrow\rangle_p$ | $\vert\uparrow\rangle_p$ |
| poles | 54($\pm$7)% | 83($\pm$6)% | 56($\pm$2)% | 73($\pm$2)% |
| equator | 66($\pm$8)% | 59($\pm$7)% | 60($\pm$3)% | 56($\pm$2)% |

**photon-to-atom swap**

|  | heralded | | unheralded | |
|---|---|---|---|---|
|  | $\vert\downarrow\rangle_a$ | $\vert\uparrow\rangle_a$ | $\vert\downarrow\rangle_a$ | $\vert\uparrow\rangle_a$ |
| poles | 63($\pm$9)% | 81($\pm$11)% | 62($\pm$4)% | 75($\pm$5)% |
| equator | 67($\pm$13)% | 84($\pm$12)% | 62($\pm$4)% | 80($\pm$5)% |

**Extended Data Table 1.** Results for the atom-to-photon, double-swap and photon-to-atom processes, broken down to "pole" and "equator" states. $\vert\uparrow\rangle_p$ and $\vert\downarrow\rangle_p$ on the photonic Bloch sphere correspond to transmission and reflection, $\vert\uparrow\rangle_a$ and $\vert\downarrow\rangle_a$ on the atomic Bloch sphere correspond to "no toggle" and "toggle", respectively. All single-swap results have been corrected to take atomic preparation errors into account.

**Acknowledgments** Support from the Israeli Science Foundation and the Crown Photonics Center is acknowledged. This research was made possible in part by the historic generosity of the Harold Perlman family.

**Author Contributions** All authors contributed to the carrying out of the experiment, discussed the results and commented on the manuscript. O.B. and A.B. analyzed the data. O.B., A.B., S.R. and B.D. contributed to the design and construction of the experimental setup. O.B., A.B., S.R. and B.D. wrote the manuscript. O.B., A.B. and S.R. contributed equally to this work.

**Author Information** The authors declare no competing financial interests. Correspondence and requests for materials should be addressed to B.D. (barak.dayan@weizmann.ac.il).